\documentclass[amssymb,12pt]{article}
\usepackage{jheppub}
\usepackage{amsmath}
\usepackage[utf8]{inputenc}
\usepackage{amsfonts}
\usepackage{amsthm}
\usepackage{graphicx}
\usepackage{slashed}
\usepackage{xcolor}
\usepackage{float}
\usepackage{slashed}
\usepackage{bm}
\usepackage[T1]{fontenc}
\usepackage{lmodern}

\usepackage{footnote}
\usepackage[bottom]{footmisc}
\newcommand{\s}{\sigma}

\newcommand{\dA}{\dot{A}}
\newcommand{\dB}{\dot{B}}
\newcommand{\dC}{\dot{C}}
\newcommand{\dD}{\dot{D}}
\newcommand{\DO}{\dot{1}}
\newcommand{\DT}{\dot{2}}

\newcommand{\xh}{\hat x}
\newcommand{\Ph}{\hat P}
\newcommand{\rh}{\hat r}
\newcommand{\that}{\hat \tau}
\newcommand{\vh}{\hat v}

\newcommand{\bz}{\bar{z}}

\newcommand{\bef}{\bar{f}}

\DeclareMathOperator{\bldF}{\mathbf{F}}
\usepackage{cancel}
\newcommand{\calO}{\mathcal{O}}

\begin{document}
\title{From Navier-Stokes to Maxwell via Einstein}
\author{Cynthia Keeler, Tucker Manton \& Nikhil Monga}
\emailAdd{nikhil.monga@asu.edu, keelerc@asu.edu \& tucker.manton@asu.edu}
\affiliation{Physics Department,
 Arizona State University, Tempe, AZ 85287, USA}
\date{May 2019}

\abstract{We revisit the cutoff surface formulation of fluid-gravity duality in the context of the classical double copy.  The spacetimes in this fluid-gravity duality are algebraically special, with Petrov type II when the spacetime is four dimensional.  We find two special classes of fluids whose dual spacetimes exhibit higher algebraic speciality: constant vorticity flows have type D gravity duals, while potential flows map to type N spacetimes.  Using the Weyl version of the classical double copy, we construct associated single-copy gauge fields for both cases, finding that constant vorticity fluids map to a solenoid gauge field.  Additionally we find the scalar in a potential flow fluid maps to the zeroth copy scalar.}

\maketitle

\section{Introduction}

The fluctuations of spacetime near a horizon can be described by a fluid equation, as first found almost forty years ago \cite{Damour:1978cg,Damour:1979wya}.  Further development of this idea led to the membrane paradigm \cite{Thorne:1986iy,Parikh:1997ma,Eling:2009sj,Eling:2009pb,Gourgoulhon:2005ng,Gourgoulhon:2005ch,Gourgoulhon:2008pu}, in which the fluid lives on a stretched horizon.  The advent of AdS-CFT duality twenty years ago allowed for a version of fluid-gravity duality where the dual fluid arises from the gauge theory living on the AdS boundary \cite{Policastro:2001yc,Policastro:2002se,Kovtun:2003wp,Kovtun:2004de,Son:2007vk,Bhattacharyya:2008kq,Iqbal:2008by,Oz:2010wz,Faulkner:2010jy,Eling:2013sna}; for reviews see \cite{Son:2007vk,Damour:2008ji,Rangamani:2009xk,Padmanabhan:2009vy,Hubeny:2011hd}.

More recently, the cutoff surface approach to fluid-gravity duality, pioneered in \cite{Bredberg:2010ky,Bredberg:2011jq} and extended in \cite{Brattan:2011my,Compere:2011dx,Compere:2012mt,Taylor:2018xcy,Bredberg:2011jq,Lysov:2011xx,Pinzani-Fokeeva:2014cka,De:2019wok,Dey:2020ogs}, built a precise version of the membrane paradigm which defines the fluid via the extrinsic curvature of an intrinsically flat hyperbolic `cutoff' surface held outside the horizon.  In the formulation of cutoff surface fluid-gravity we follow in this paper, \cite{Bredberg:2011jq}, the Einstein constraint equations on the hyperbolic cutoff surface become the nonlinear incompressible Navier-Stokes equations, while solving the remaining Einstein equations defines the rest of the spacetime.  We will work mostly with the low order terms in the long-wavelength or hydrodynamic limit, which amounts to a gradient expansion; as shown in \cite{Compere:2011dx}, this procedure does allow a full perturbative expansion.

The classical double copy as first presented in \cite{Monteiro:2014cda} builds a map between classical gravity solutions and classical Yang-Mills solutions, based on the color-kinematics duality valid at the amplitude level (see \cite{Bern:2019prr} for a comprehensive review). Since the metric of the gravitational solution is built out of two copies of the classical Yang-Mills solution, the Yang-Mills solution is referred to as the `single copy' of the corresponding metric, and there is also a corresponding Klein-Gordon scalar solution termed the `zeroth copy'. As an example, the single copy of the Schwarzschild black hole metric is the field arrangement due to a color charge at the origin, when the dilaton expectation value is tuned to zero \cite{Luna:2020adi}.  Many other examples of the classical double copy have been built \cite{Luna:2015paa,Luna:2016hge,Monteiro:2018xev,Alfonsi:2020lub,Ridgway:2015fdl,Adamo:2017nia,Adamo:2017nia,Bahjat-Abbas:2017htu,Carrillo-Gonzalez:2017iyj,Ilderton:2018lsf,Gurses:2018ckx,CarrilloGonzalez:2019gof,Lee:2018gxc,Bah:2019sda,Andrzejewski:2019hub,Goldberger:2019xef,Kim:2019jwm,Bahjat-Abbas:2020cyb}, including to some broad classes of spacetime \cite{Luna:2018dpt}.  Furthermore \cite{Goldberger:2017frp,Luna:2017dtq,Goldberger:2017vcg,Shen:2018ebu,Cheung:2018wkq,Kosower:2018adc,Bern:2019nnu,Antonelli:2019ytb,Bern:2019crd,Kalin:2019rwq} have used this classical mapping to improve the perturbative series used in analytic calculations of black hole collisions.

We build herein the single copy gauge fields which map to fluid-dual metrics, for two different classes of Navier-Stokes solutions.  We are able to accomplish this map by relying on the algebraic speciality of these fluid-dual metrics.  A spacetime is algebraically special if its Weyl tensor exhibits extra symmetry; specifically, if two or more of its principal null vectors coincide. In four dimensions, spacetimes of Petrov type D have two pairs of coinciding principle null vectors, while spacetimes of type N have all four principal null vectors coincident. Using the constrained form of the Weyl tensor for algebraically special spacetimes, \cite{Luna:2018dpt} exhibited a single copy gauge field (and zeroth copy scalar field) valid for every type D vacuum solution to general relativity.  

As \cite{Bredberg:2011jq,Lysov:2011xx} note, the spacetime corresponding to the fluid metric is algebraically special; for four dimensions, the spacetime has Petrov type II. As we will show, further restricting the fluid results in higher algebraic speciality.  We focus on two special fluid classes: constant vorticity fluids and potential flows.  Constant vorticity fluids are dual to spacetime metrics with Petrov type D, while potential flow fluids are dual to metrics with Petrov type N.  Such fluids have also been studied in the context of holography, for instance for flows with vorticity \cite{Leigh:2011au,Leigh:2012jv}. Consequently, using the Weyl double copy proposed in \cite{Luna:2018dpt}, we are able to exhibit the single copy gauge fields whose double copy metric is then dual to either a constant vorticity fluid or a potential flow fluid.  Since these these gauge fields are in the U(1) sector of the Yang-Mills theory, we have thus mapped two classes of Navier-Stokes solutions to Maxwell solutions.

The gauge field corresponding to the constant vorticity fluid matches the constant axial field within a large solenoid, while the zeroth copy is a constant.  For the potential flow fluids, the gauge field is the same for every potential flow; it corresponds to a static Maxwell field with Poynting vector pointing towards the horizon. We find the scalar flow potential maps to the zeroth copy scalar field.  Thus, just as the nontrivial details of the constant vorticity fluid map to the single copy field, the nontrivial details of the potential flow fluid map instead on to the zeroth copy scalar potential.

In section \ref{fluidgravintro} we begin by reviewing the cutoff approach to fluid-gravity duality from \cite{Bredberg:2011jq}. In section \ref{classicaldc}, we briefly review the classical double copy story, focussing on the Weyl double copy as developed in \cite{Luna:2018dpt}. In section \ref{fluidsolutionsSection} we show that constant vorticity fluids map to type D vacuum metrics, while potential flow fluids map to type N metrics. In sections \ref{type-D} and \ref{type-N} we build the single copy for the gauge fields associated with these metrics. In section \ref{discussions} we discuss the physical implications of our results and speculate on the viability of a classical double copy picture for generic fluid-dual spacetimes.

\section{The Hydrodynamic Limit and Near-Horizon Expansion}\label{fluidgravintro}

In this section we review the cutoff surface formulation of fluid-gravity duality and reiterate the equivalence between the hydrodynamic limit and the near horizon expansion explored in \cite{Bredberg:2011jq}.  In order to obtain Navier-Stokes equations from Einstein's equations, we begin with a background Rindler spacetime written in ingoing Eddington-Finkelstein coordinates:
\begin{equation}
ds^2_0=-r d\tau^2 + 2 d\tau dr + dx_i dx^i.
\end{equation}
Here $i,j$ will be the spacelike fluid directions; for a fluid in $2+1$ dimensions, $i,j$ run over $1,2$ and the associated metric is four-dimensional. Constant $r$ hypersurfaces in these coordinates are intrinsically flat and foliate the spacetime metric into hyperbolic slices.  

We then choose one such slice, $r=r_c$, and perturb the spacetime there, generating extrinsic curvature for the $r=r_c$ slice as embedded in the full spacetime.  We identify this extrinsic curvature $\kappa_{ab}$ with the fluid stress tensor $T_{ab}$; here $a, b$ run over the directions along the $r_c$ slice (that is, $a,b$ take values $\tau$ or $i,j$).  The intrinsic metric of this slice $\gamma_{ab}$ thus satisfies 
\begin{equation}
    \gamma_{ab} = -r_c d\tau^2 +dx^jdx_j,  \qquad \gamma_{ab}\kappa-\kappa_{ab} \sim T^{NS}_{ab}.
\end{equation}
For these perturbations, we impose regularity and infalling boundary conditions at the null horizon $r=0$, thus generating the fluid-dual metric
\begin{equation}
\label{mrc} 
\begin{split}
    ds^2 =&-r d\tau^2 + 2 d\tau dr + dx_i dx^i   \\
    &-2\left(1-\frac{r}{r_c}\right) v_i dx^i d\tau -2\frac{v_i}{r_c} dx^i dr  \\
    &+{\left(1-\frac{r}{r_c}\right)}\left[ (v^2+2P) d\tau^2  +\frac{v_i v_j}{r_c} dx^i dx^j \right] + \left( \frac{v^2}{r_c} +\frac{2P}{r_c}\right) d\tau dr     \\
    &-{\frac{(r^2 - r_c^2)}{r_c}}\partial^2 v_i dx^i d\tau + \calO(\epsilon^3).
\end{split}
\end{equation}
The $\epsilon$ here refers to the order in the hydrodynamic or long wavelength expansion, explicitly 
\begin{equation}
\label{hydrolimit}
\partial_i\to\epsilon, \qquad \partial_\tau \to \epsilon^2, \qquad v \to \epsilon, \qquad P \to \epsilon^2.
\end{equation}
The metric in \eqref{mrc} is arranged with background terms of order $\calO(\epsilon^0)$ in the first line, $\calO(\epsilon)$ terms in the second, and so on.

With these identifications, the $r=r_c$ constraint components of Einstein's equations, $G_{\tau \tau}$ and $G_{\tau i}$, become incompressibility and the Navier-Stokes equation:
\begin{equation}
\begin{split}
    G_{00} &= 0 \implies \partial_i v_i = 0, \\
    G_{0i} &= 0 \implies \partial_\tau v_i-\eta\partial^2 v_i+\partial_i P+v^j\partial_j v_i=0,
\end{split}    
\end{equation}
where the shear viscosity $\eta$ is identified%
\footnote{Note that in the near horizon expansion $\eta\to1$.}
with $r_c$. 

As in \cite{Bredberg:2011jq}, to relate the hydrodynamic limit to the near horizon limit, we introduce hatted coordinates and variables: 
\begin{equation}
\label{hattedcoords}
    x_i = \frac{r_c \hat{x}_i}{\epsilon}, \qquad \tau = \frac{r_c \hat{\tau}}{\epsilon^2}, \qquad r = \hat{r}r_c ,\qquad v_i = \epsilon \hat{v}_i \qquad P = \epsilon^2 \hat{P}.
\end{equation}
Next, we rescale the metric and define a new perturbative parameter $\lambda$:
\begin{equation}
\label{rescale}
     ds^2 \to d\hat{s}^2=\frac{\epsilon^2}{r_c^2} ds^2 \qquad z^2 - t^2 = 4 r_c \to  4 \lambda, \qquad \lambda \equiv \frac{\epsilon^2}{r_c} .
\end{equation}
This new expansion parameter $\lambda$ controls the near horizon expansion. The limit $\lambda\to 0$ sets the $r=r_c$ hypersurface to be null, just like the $r=0$ Rindler horizon. In the near horizon expansion the metric thus becomes
\begin{equation}
\begin{split}
\label{lambdaexpansion}
d\hat{s}^2 =&-\frac{ \rh  }{\lambda} d \that ^2 \\ & + \left[2 d\that d\rh + d\xh_i d\xh^i -2 (1-\rh) \vh_i d \xh^i d\that +{(1-\rh)}(\vh^2+2\Ph) d\that^2  \right]
\\
& +\lambda\big[{(1-\rh)}{\vh_i \vh_j }d\xh^i d\xh^j -2{\vh_i}d\xh^i d\rh   + (\vh^2+2\Ph)d\that d\rh 
\\
&+
(\rh-1)[-(\rh+1) \hat\partial^2 \vh_i + (\vh^2+2\Ph)2\vh_i +4 \hat\partial_i\Ph]d\xh d\that\big] + \calO(\lambda^2).
\end{split}
\end{equation}
In this sense \cite{Bredberg:2011jq} demonstrate that the near horizon expansion matches the long wavelength limit, consistent with the perspective that horizons behave as incompressible fluids. 

As discussed further in appendix \ref{TetradsAppendix}, the replacements 
\begin{equation}
\label{hydroreplace}
x^i\to\epsilon x^i, \qquad \tau \to \epsilon^2\tau , \qquad v \to \epsilon v, \qquad P \to \epsilon^2 P.
\end{equation}
allow derivation of the incompressible Navier-Stokes equation starting from a solution of more complicated equations; essentially, any other terms become higher order terms in the $\epsilon$ expansion. Additionally, these replacements will bring a Navier-Stokes solution that is not initially in the long wavelength limit  \eqref{hydrolimit} into that limit.  The near horizon expansion makes these replacements explicit, so it is valid for Navier-Stokes solutions that are not naturally in the hydrodynamic limit, such as vortices.  Consequently, although we mostly use the hydrodynamic expansion $\epsilon$ below, we will return to the near horizon $\lambda$ expansion when necessary.

\section{Classical Double Copy}\label{classicaldc}

In the past few decades, significant steps have been made towards a deeper understanding of graviton scattering amplitudes and their relation to gauge scattering amplitudes. Most relevant for this article is the double copy prescription (see \cite{Bern:2019prr} and references within for a comprehensive review of the subject). Stated simply, the double copy obtains complicated graviton scattering amplitudes from simpler gauge theory amplitudes. The gauge theory amplitude $\mathcal{A}^{\text{YM}}$ is written in a generalized gauge such that it takes the schematic form
\begin{equation}\label{gaugeamp}
\mathcal{A}^{\text{YM}}\sim\sum_k \frac{n_k c_k}{\text{propagators}},
\end{equation}
where the sum is over all three-point vertex graphs, the $n_k$ are the kinematic numerators associated with each graph, and the $c_k$ are the color factors that satisfy a Jacobi identity of the form $c_i+c_j+c_k=0$. The basic principle in obtaining the graviton amplitude relies on a particularly simple duality between color and kinematics, the BCJ duality first presented in \cite{Bern:2008qj}, being made manifest. 

The double copy prescription then provides the corresponding graviton amplitude,
\begin{equation}\label{gravamp}
\mathcal{M}^{\text{grav}}\sim \sum_k \frac{n_k n_k}{\text{propagators}},
\end{equation}
where the color factors $c_k$ have been replaced with a second set of kinematic numerators $n_k$ that are organized to also satisfy a Jacobi identity of the same form. There is also a `zeroth copy' in the amplitudes story, where starting with (\ref{gaugeamp}), replacing the kinematic numerators $n_i$ with a second set of color factors $\tilde{c}_i$ builds scalar amplitudes of the form
\begin{equation}\label{scalaramp}
\mathcal{A}^{\text{scalar}}\sim\sum_k\frac{c_k\tilde{c}_{k}}{\text{propagators}},
\end{equation}
for bi-adjoint scalars $\phi^{aa'}$. As we will see below, a zeroth copy scalar can also be found in the classical double copy story; it will play a significant role for the potential flow fluid class.

When the double copy procedure is applied to pure (non-supersymmetric) Yang-Mills theory, the resulting theory on the gravity side is general relativity coupled to a two-form field and a dilaton. Although these amplitude relations are perturbative quantum statements, the authors of \cite{Monteiro:2014cda} used these relations to inspire a double copy mapping between classical solutions in general relativity and classical solutions in the $U(1)$ sector of Yang-Mills.%
\footnote{Some nonabelian behavior is covered in e.g. \cite{Alfonsi:2020lub,Bahjat-Abbas:2020cyb}, but here we focus on only the abelian sector.}
This relation is referred to as the classical double copy. 

\subsection{Kerr-Schild Double Copy}

The key connection between the classical gravity and gauge theory solutions first presented in \cite{Monteiro:2014cda} is the use of Kerr-Schild coordinates, where
\begin{equation}\label{KerrSchild}
g_{\mu\nu}=\eta_{\mu\nu}+\phi k_\mu k_\nu.
\end{equation}
Here, $\phi$ is a scalar function that plays the role of the zeroth copy, and satisfies the wave equation over the flat background, $\eta^{\mu\nu}\partial_\mu\partial_\nu\phi=0$. The vector $k_\mu$ is null with respect to both the full and background metrics,
\begin{equation}
g^{\mu\nu}k_\mu k_\nu = \eta^{\mu\nu}k_\mu k_\nu = 0. 
\end{equation}
This feature serves to truncate the inverse metric to $g^{\mu\nu}=\eta^{\mu\nu}-\phi k^\mu k^\nu,$  with the further consequence that the null vector can be raised with either the background or full metric, $k^\mu=g^{\mu\nu}k_\nu=\eta^{\mu\nu}k_\nu$.

The classical double copy states that if $g_{\mu\nu}$ is a solution to the Einstein equations, then the gauge field given by
\begin{equation}\label{gaugefield}
A_\mu^a = c^a\phi k_\mu 
\end{equation}
is a solution to Yang-Mills theory.  Since the $c^a$ are just constant color factors in these solutions, these solutions really live in a $U(1)$ sector of the gauge theory; that is, $A_\mu=\phi k_\mu$ will be a Maxwell solution. We refer to (\ref{gaugefield}) as the single copy, in line with terminology in the amplitudes story.

 The connection between the classical story and amplitudes story can be seen by replacing the color vector $c^a$ in (\ref{gaugefield}) with the null vector $k_\mu$ in (\ref{KerrSchild}) to obtain $h_{\mu\nu}$ from the gauge theory, akin to replacing $c_k\rightarrow n_k$. Moreover, the zeroth copy analogy can be seen by replacing $k_\mu\rightarrow c^{a'}$ in (\ref{gaugefield}) to get $\phi^{aa'}=c^a c^{a'}\phi$, in the same spirit as replacing $n_i\rightarrow \tilde{c}_{i}$ to obtain (\ref{scalaramp}) from (\ref{gaugeamp}). The mapping (\ref{gaugefield}) has been extensively studied for various exact solutions living on flat space \cite{Monteiro:2014cda,Adamo:2017nia,Luna:2015paa,Goldberger:2017vcg,CarrilloGonzalez:2019gof,Ridgway:2015fdl,Luna:2016due,Goldberger:2016iau,Berman:2018hwd,Gurses:2018ckx,PV:2019uuv,Ilderton:2018lsf} and extended to solutions living on maximally-symmetric backgrounds \cite{Carrillo-Gonzalez:2017iyj,Bahjat-Abbas:2017htu}.
 
Some classical solutions that have been shown to exhibit a reasonable double copy necessitate an extension to the ansatz (\ref{KerrSchild}); \cite{Luna:2015paa,Luna:2018dpt,Lee:2018gxc} write the full metric in double Kerr-Schild form, where 
\begin{equation}\label{doubleKS}
g_{\mu\nu}=\eta_{\mu\nu}+\phi k_\mu k_\nu+\psi l_\mu l_\nu.
\end{equation}
Here the vectors $k$ and $l$ are individually null as well as orthogonal (orthonullity);
\begin{equation}\label{orthonull}
k^2=l^2=k\cdot l = 0.
\end{equation}
Again, the indices for both vectors can be raised and lowered with either the full metric $g_{\mu\nu}$ or the background metric $\eta_{\mu\nu}$. This form was necessary for the single copy study of the Taub-NUT solution \cite{Luna:2015paa} as well as for the generic type D vacuum solutions in \cite{Luna:2018dpt}, where the gauge field is given by 
\begin{equation}
A^a_\mu = c^a \big(\phi k_\mu+\psi l_\mu\big).
\end{equation}
\subsection{Weyl Double Copy}

In our work, we will utilize a different realization of the classical double copy, referred to as the Weyl double copy \cite{Luna:2018dpt}. This prescription for the double copy relies on the spinor formulation of general relativity \cite{Penrose:1987uia,Penrose:1986ca} in conjunction with the Petrov classification (see \cite{Stephani:2003tm} chapters 3 and 4 for a review of both concepts) to build the map between the gravitational and gauge theories. This version of the double copy applies to four-dimensional spacetimes, although \cite{Monteiro:2018xev} builds towards an extension to higher dimensions; for now we review the four-dimensional picture.

The Petrov classification labels metrics by the multiplicities of the principle null directions of their Weyl tensors.  A principle null direction $k^\mu$ satisfies
\begin{equation}
k_\mu k^\mu=0, \qquad k_{[\sigma}W_{\mu]\nu\rho[\sigma}k_{\lambda]}k^{\nu}k^{\rho}=0,
\end{equation}
where $W_{\mu\nu\lambda\gamma}$ is the Weyl tensor.  All four-dimensional metrics will have four (not necessarily unique) solutions $k^\mu$ to these equations, but they can appear with different multiplicities.  A spacetime is algebraically special if any two or more of these principle null vectors coincide.  If only two coincide, the spacetime is Petrov type II; if two pairs coincide, then it is type D.  If all four principle null vectors coincide, then the spacetime is type N.  The Weyl double copy will apply to type D and type N spacetimes, essentially factoring their principle null vector pairs.

Since a basic understanding of curved space spinor formalism is necessary to work with the Weyl double copy, we review the essentials in appendix \ref{spinorappendix}. We rewrite the usual Weyl tensor $W_{\mu\nu\lambda\gamma}$ in terms of the completely symmetric Weyl spinor $C_{ABCD}$ using the formula
\begin{equation}
C_{ABCD}=\frac{1}{4}W_{\mu\nu\lambda\gamma}\sigma^{\mu\nu}_{AB}\sigma^{\lambda\gamma}_{CD},
\end{equation}
where $\sigma^{\mu\nu}_{AB}$ are defined in terms of the Pauli sigma matrices as in (\ref{bigsigmas}).

The form of the Weyl spinor $C_{ABCD}$ is directly related to the Petrov classification of spacetimes, since the Weyl spinor can be decomposed as  
\begin{equation}
C_{ABCD}=\alpha_{(A}\beta_B \gamma_C \delta_{D)},
\end{equation}
where the four principle spinors $\{\alpha_A, \beta_B, \gamma_C, \delta_D\}$ carry the information of the four principle null directions of the spacetime. The principle spinors can be related to the principle null vectors using the Pauli 4-vectors via \eqref{omicronsANDiotas}.

Since the spinors composing $C_{ABCD}$ are directly related to the principle null vectors, their multiplicity also depends on the Petrov type. If all four spinors are unique, the spacetime is algebraically general, of Petrov type I. Otherwise the spacetime is algebraically special. We focus on Petrov type D, where there are two unique principle spinors with multiplicity two, and Petrov type N, where there is one unique principle spinor.  Their Weyl spinors can be written 
\begin{equation}
C_{ABCD}^{\text{D}}\sim \alpha_{(A}\alpha_B \beta_C \beta_{D)}, \qquad C_{ABCD}^{\text{N}}\sim\alpha_A \alpha_B \alpha_C \alpha_D,
\end{equation}
where here $\alpha$ (and $\beta$, for type D) are the principle null spinors. 

On the gauge theory side, the spinor field strength $f_{AB}$ is the key object, and can be obtained from the field strength tensor $F_{\mu\nu}$ directly using 
\begin{equation}
f_{AB}=\frac{1}{2}F_{\mu\nu}\sigma^{\mu\nu}_{AB}.
\end{equation}
In the same sense as the Weyl spinor, the $f_{AB}$ corresponding to a type D spacetime can be written as $f^{\text{D}}_{AB}\sim \alpha_{(A}\beta_{B)}$, whereas in the type N case we have $f^{\text{N}}_{AB}\sim \alpha_A \alpha_B$. Thus we find
\begin{equation}\label{WeylDCrelation}
C_{ABCD}=\frac{1}{S}f_{(AB}f_{CD)},
\end{equation}
where $S$ is a complex scalar field satisfying the wave equation in the flat background on which $f_{AB}$ lives, and whose real part coincides with the Kerr-Schild scalar $\phi$ up to an overall constant. Therefore the scalar $S$ plays the role of the zeroth copy in the Weyl double copy map.

We will use the decomposition of the Weyl spinor $C_{ABCD}$ in terms of a spinor basis $\{o_A,\iota_B\}$:
\begin{equation}\label{expandCABCD}
    C_{ABCD}=\Psi_0 \iota_{A}\iota_B\iota_C\iota_D-4\Psi_1 o_{(A}\iota_B\iota_C\iota_{D)}+6\Psi_2 o_{(A}o_B\iota_C\iota_{D)}-4\Psi_3 o_{(A}o_Bo_C\iota_{D)}+\Psi_4 o_Ao_Bo_Co_D.
\end{equation}
Here, the $\Psi_I\in \mathbb{C}, \ I=0,1,2,3,4$ are called Weyl scalars, and are also related to the Petrov classification (see section \ref{fluidsolutionsSection}). We will see that the $\Psi_I$, and the invariants built out of them, play a significant role in the Weyl double copy.

As \cite{Luna:2018dpt} shows, solutions built from this Weyl double copy picture match the expectations from the Kerr-Schild double copy as built in \cite{Monteiro:2014cda}.  In addition to specific examples like the Kerr metric, \cite{Luna:2018dpt} also shows this matching for the most general type D vacuum solution as written in Plebanski-Demianski coordinates \cite{Plebanski:1976gy} (see \cite{Griffiths:2005qp} and \cite{Podolsky:2018nkx} for an extended treatment). 

We next look to analyze solutions to Navier-Stokes from the fluid gravity perspective that result in spacetimes that are candidates for the Weyl double copy. As we will now show, by constraining the velocity fields in the fluid metric (\ref{mrc}) in one of two ways, we find that the resulting spacetime is either Petrov type N or type D, allowing for a double copy treatment via the Weyl method.

\section{Fluid Solutions}\label{fluidsolutionsSection}

The eigenbivectors of the Weyl tensor for the fluid metric reveal that it is algebraically special \cite{Bredberg:2011jq,Lysov:2011xx}; specifically it is a type II spacetime according to the Petrov classification, with two coinciding principal null vectors. Below, we use the Newman-Penrose formalism to find which fluids correspond to metrics with even higher algebraic speciality. Additional details pertaining to the formalism and our choice of conventions can be found in Appendix \ref{NPappendix} or in  \cite{Stephani:2003tm}.

Briefly, the Newman-Penrose formalism relies on rewriting the metric in terms of a tetrad set $l,\, n,\, m,\, \bar m$, as in \eqref{tetradmetric}. The tetrad set is then used to compute the Weyl scalars, which then can be used to compute the invariants $I,\,J,\, K,\, L,$ and $N$ as in \eqref{Weylinvariants}. While the Weyl scalars depend on the tetrad choice, the invariants do not and thus we will look at these invariants to classify our spacetimes. 

We work in the hydrodynamic limit of the metric \eqref{mrc}, where the first terms we do not write explicitly%
\footnote{ \cite{Lysov:2011xx} show that algebraically special spacetimes can be obtained to arbitrary order in the context of the fluid gravity duality in 5 or higher spacetime dimensions. \cite{Cai:2013uye} also consider similar spacetimes in $d \geq 5$, however posit that additional constraints may be needed in \cite{Lysov:2011xx} at higher orders to maintain algebraic speciailty. \cite{Compere:2011dx} construct a formulation that progresses to arbitrary order, however this construction deviates from algebraic speciality and in doing so relates the higher order pieces in the metric to corrections to the Navier-Stokes equations. Since our interest is primarily in making connection with the Weyl double copy picture, we restrict ourselves to the first few nontrivial orders of this metric. For more on convergence of the gradient expansion in a hydrodynamic and fluid gravity context, see \cite{Pinzani-Fokeeva:2014cka,Grozdanov:2019kge}.} 
arise at $\calO(\epsilon^3)$.  Thus we only know our Weyl scalars up to the same order, and our algebraic classification of the spacetime is perturbative as well.
In this limit, our tetrad choice \eqref{epsilontetrad} yields the following Weyl scalars up to $\calO(\epsilon^3)$, which is where we would start to see contributions from neglected higher terms in the metric\eqref{mrc}:
\begin{align}
\label{epsilonWeylscalars}
   \Psi_{0} &= 0 + \calO(\epsilon^{3}),     \notag \\
   \Psi_{1} &= 0 + \calO(\epsilon^{3}), \notag\\
   \Psi_{2} &=  -i\frac{\epsilon^2}{4r_c}(\partial_x v_y - \partial_y v_x) + \calO(\epsilon^{3}),\\
   \Psi_{3} &= 0 + \calO(\epsilon^{3}), \notag \\
    \Psi_{4} &= - \frac{\epsilon^2}{2r}(\partial_x v_x - \partial_y v_y + i (\partial_x v_y + \partial_y v_x))+ \calO(\epsilon^{3}). \notag
\end{align}
$\Psi_2$ is proportional to the vorticity of the fluid, while $\Psi_4$ is proportional to the derivative of $v_x+i v_y$ with respect to the complex coordinate $\bar{z}\equiv x -iy$.  

In order to evaluate the algebraic speciality of our spacetimes, we compute the invariants $I, \ J, \ K, \ L$ and $N,$ via the following relations:
\begin{equation}\label{Weylinvariants}
    \begin{split}
        I&\equiv\Psi_0\Psi_4-4\Psi_{1}\Psi_{3}+3\Psi_{2}^{2}, \\
        J&\equiv \begin{vmatrix} 
        \Psi_{4}&\Psi_{3}&\Psi_{2}\\
        \Psi_{3}&\Psi_2 &\Psi_1 \\
        \Psi_{2}&\Psi_{1} &\Psi_{0}
        \end{vmatrix} ,\\
        K&\equiv \Psi_{1}\Psi_{4}^{2}-3\Psi_{4}\Psi_{3}\Psi_{2}+2\Psi_{3}^{3}, \\
        L&\equiv \Psi_{2}\Psi_{4}-\Psi_{3}^{2}, \\
        N&\equiv 12L^{2}-\Psi_{4}^{2}I.
    \end{split}
\end{equation}
For a generic fluid-dual metric, we find 
\begin{equation}\label{IJ}
\begin{split}
  I &= 3 \epsilon ^4 \left[ \,i\,\left(\frac{ \partial_x
   v_y}{4r_c}-\frac{ \partial_y v_x}{4 r_c}\right)\right]^2 + \calO(\epsilon^{5}),\\
  J &= \epsilon ^6\left[ \,i\,\left(\frac{ \partial_x
   v_y}{4r_c}-\frac{ \partial_y v_x}{4 r_c}\right)\right]^3+ \calO(\epsilon^{7}).\\
\end{split}
\end{equation}
These $I$ and $J$ satisfy $I^3-27 J^2 = 0$, or more precisely,
\begin{equation}
   \implies I^3 - 27 J^2 = 0 + \calO(\epsilon^{13}),
\end{equation}
which implies that the general fluid metric is Petrov type II up to this order. 

Next we look at the invariants $K, \, L,$ and $N$:
\begin{equation}\label{KLN}
\begin{split}
    K &= 0 + \calO(\epsilon^{7}),\\
    L &= \epsilon^4 \biggr[-\frac{\partial_x v_x}{2r}+\frac{\partial_y v_y}{2r} - i \frac{\partial_y v_x}{2r} - i \frac{\partial_x v_y}{2r}\biggr]\biggr[i\,\frac{\partial_y v_x}{4 r_c}-i\,\frac{\partial_x v_y}{4 r_c}\biggr]+ \calO(\epsilon^{5}),\\
    N &= 9 \epsilon^8 \biggr[-\frac{\partial_x v_x}{2r}+\frac{\partial_y v_y}{2r} - i \frac{\partial_y v_x}{2r} - i \frac{\partial_x v_y}{2r}\biggr]^2\biggr[i\,\frac{\partial_y v_x}{4 r_c}-i\,\frac{\partial_x v_y}{4 r_c}\biggr]^2+ \calO(\epsilon^{9}) .
\end{split}
\end{equation}
Although $K$ is in fact 0 through this order, that is not enough for further algebraic speciality (see Figure 9.1 in \cite{Stephani:2003tm}). The nonzero invariants $L$ and $N$ are proportional to both the vorticity (from $\Psi_2$) and $\partial_{\bar z}(v_x+iv_y)$ (from $\Psi_4$).

Before we begin an analysis of which special fluids have dual metrics with higher algebraic speciality, we must mention briefly the perturbative nature of the metrics we use in this paper. 
While \cite{Lysov:2011xx} constructed fluid-dual spacetimes by requiring algebraic speciality to hold at all orders, here we instead constrain ourselves only to the lowest orders necessary in order to establish the incompressible Navier-Stokes equations.  Accordingly, we only establish the higher algebraic speciality of our spacetimes to lowest order.

To these orders discussed, the condition that the fluids spacetime is a type II metric, $I^3-27J^2=0$, is satisfied in either the near-horizon or the hydrodynamic expansion:
\begin{equation}
I_{\epsilon}^3 - 27 J_{\epsilon}^2 = 0 + \calO(\epsilon^{13}),\\
 \qquad I_{\lambda}^3 - 27 J_{\lambda}^2 = 0 + \calO(\lambda).
\end{equation}
Note that the highest non-error order available in the near-horizon $\lambda$ expansion differs from the $\epsilon$ hydrodynamic expansions, but both spacetimes satisfy the type II constraint to at least one nontrivial order.  

Specifically, in the near-horizon expansion, we find 
\begin{equation}
\begin{split}
    I_\lambda &= -\frac{3}{16}(\partial_y v_x - \partial_x v_y)^2 +  {\calO(\lambda)},\\
    J_\lambda &= -\frac{i}{64}(\partial_y v_x - \partial_x v_y)^3 +  {\calO(\lambda)},
\end{split}    
\end{equation}
which matches \eqref{IJ} except for the expansion order.
Since the order of terms differs between the two expansions, in the near-horizon expansion it turns out to be necessary to account for terms of order $\calO(\lambda^2)$ in the metric \eqref{lambdaexpansion}, as was done in \cite{Bredberg:2011jq}.  Accordingly we use the generic form of the tetrad \eqref{lambdatetrad} to perform computations in this expansion.

Since the fluid constraints required to produce higher algebraic speciality are the same at the lowest order of both expansions, we thus concentrate on only the $\epsilon$ hydrodynamic expansion for the remainder of this section.  As we show below, constant vorticity fluids will correspond to type D spacetimes while potential flows correspond to type N metrics.

\subsection{Petrov Type D Fluid Solutions}\label{typeDfluidsSection}

A Petrov type D spacetime satisfies the following conditions for the invariants: 
\begin{equation}\label{typeDcond}
    I^3 - 27 J^2 = 0;  \qquad I,J \neq 0; \qquad K = N = 0. 
\end{equation}
Based on the forms of $L$ and $N$ in \eqref{KLN} and $I$ and $J$ in \eqref{IJ}, these conditions imply
\begin{equation}
     \partial_x v_y - \partial_y v_x \neq 0, \qquad   -\partial_x v_x+\partial_y v_y-i\Big(\partial_y v_x+\partial_x v_y\Big)=0.
\end{equation}
These constraints imply that each component of the velocity satisfies Laplace's equation $\partial^2 v_i = 0$, where $i\in\{x,y\}$. 

These conditions are solved by the fluid velocities
\begin{equation}\label{typeDvelocity}
\begin{split} v_x(\tau,y)&=-\omega y+h_x(\tau),\\ v_y(\tau,x)&=\omega x+h_y(\tau),\end{split}
\end{equation}
with pressures
\begin{equation}\label{typeDpressure}
 P(\tau,x,y)=\frac{\omega^2 }{2}\big(x^2+y^2\big)+\big(\omega h_y-\partial_\tau h_x\big)x-\big(\omega h_x+\partial_\tau h_y\big)y+c(\tau).
\end{equation}

In this paper, we will concentrate on the steady state solution centered at the origin; that is, we set  $h_i(\tau)=c(\tau)=0$. Turning these functions on would correspond to a vortex whose center follows the path  $(x_0(\tau),y_0(\tau))=(\int h_x d\tau,\int h_y d\tau)$ as time $\tau$ passes; a diffeomorphism returning to coordinates centered on the moving vortex would tune the effective time dependence back to zero.

Thus the fluid profile we study as representative of fluids dual to type D metrics satisfies
\begin{equation}\label{typeDvandP}
 v_x(\tau,y)=-\omega y,\qquad v_y(\tau,x)=\omega x,\qquad P = \omega^2\frac{ (x^2+y^2)}{2},
\end{equation}
consistent with vanishing pressure and velocity at the origin as would be expected for a fluid rotating with constant vorticity, centered at the origin. 
\subsection{Petrov Type N Fluid Solutions}\label{typeNfluids}

To obtain a type N spacetime, the invariants must satisfy
\begin{equation}
I=0,\quad J=0, \quad K =0, \quad L = 0, \quad N\neq0.
\end{equation}
For the general fluid metric, we already have $K=0$ and the invariants $I,J$ \eqref{IJ} and $L$ \eqref{KLN} are each proportional to a positive power of the vorticity, so setting the fluid vorticity $\partial_x v_y - \partial_y v_x $ to zero leaves us with a type N dual metric.

The velocity and pressure profiles of vorticity-free fluids can be written in terms of a scalar potential $\phi$:
\begin{equation}\label{fluidpotflow}
 v_i = \partial_i \phi, \qquad \partial_i P = - \partial_i\partial_{\tau} \phi  - \partial^j \phi \partial_i \partial_j \phi.
\end{equation}
For incompressible fluids, $\phi$ satisfies Laplace's equation $\partial^2\phi=\partial_x^2\phi+\partial_y^2\phi=0$, so vorticity-free incompressible fluids are referred to as potential flows.  

These potential flows can be written cleanly in complex coordinates, i.e. using $z\equiv x+i y$. Since $\partial^2 \phi=0$, we can rewrite a general solution for the potential $\phi$ using the sum of a holomorphic function $f$ and an antiholomorphic function $g$:
\begin{equation}\label{holomorphicPotFlow}
\partial_{z}\partial_{\bar{z}} \phi = 0, \qquad \phi = f(z) + g(\bar{z}).
\end{equation}
Imposing reality conditions so as to obtain real velocity and pressure fields requires that the antiholomorphic function $g(z)$ must be the complex conjugate of the function $f(z)$: 
\begin{equation}\label{holomorphicPotFlowB}
 \phi =  f(z) +\bar{f}(\bar{z}), \qquad \bar{f}(\bar{z})\equiv (f(z))^*.
\end{equation}

Returning to the dual fluid metric, the vorticity-free condition sets $\Psi_2=0$, leaving only $\Psi_4$ nonzero. We can express this nonzero Weyl scalar compactly as
\begin{equation}\label{psi4}
        \Psi_4 = -\frac{2}{r}\partial_{\bar{z}}^2\phi= - \frac{2}{r}\partial_{\bar{z}}^2 \bar{f}(\bar{z}),
\end{equation}

while the Weyl tensor becomes 
\begin{equation}\label{typeNWeyl}
 C_{ABCD} = \Psi_4 o_A o_B o_C o_D.
\end{equation}

Since the function $f(z)$ is holomorphic, we can write a general fluid solution as a Laurent series in $z$ (and $\bar{z}$ for $\bar f$): 
\begin{equation}\label{AlphaPot}
\phi = \sum_{n={-\infty}}^{\infty}\alpha_{n+2} z^{n+2} + \text{c.c}.,  
\end{equation}
where $\alpha_n$ are in general complex valued coefficients and the holomorphic function $f(z) \equiv \sum_{n={-\infty}}^{\infty}\alpha_{n+2} z^{n+2}$. Consequently the Weyl scalar $\Psi_4$ can also be written as a Laurent series.

It is instructive to look at the forms of the fluid potential and the Weyl scalars for a few specific fluid solutions here%
\footnote{Note as for the type D case, we neglect the time dependence that could be allowed in the $\alpha$ coefficients of the fluid potential and instead consider only steady state flows. As before, time dependence here will correspond to translating these steady state solutions.}%
. We begin by turning on only the $n=0$ term in \eqref{AlphaPot}.  For convenience we additionally choose $\alpha_{2}=-\alpha/4$, with $\alpha$ real, obtaining the potential 
\begin{equation}
\phi(z,\bar{z})  =  -\frac{\alpha}{4} (z^2+\bar{z}^2). 
\end{equation}
The corresponding fluid velocity and pressure profiles become
\begin{equation}\label{TypeNfluidSolutions}
    v_x = - \alpha x,  \qquad v_y = \alpha y,  \qquad P = P_0 - \alpha^2\,\, \frac{x^2+y^2}{2}.
\end{equation}
This fluid profile is known as planar extensional flow; extensional flows have been well studied in the fluid-mechanics/materials science community, see e.g. \cite{barnes1989introduction}.  Our main interest in this fluid will be its simplicity in terms of the double copy prescription, as we will see below.
 
Using \eqref{psi4}, for this fluid we find
\begin{equation}
        \Psi_4 = \frac{\alpha }{ r}.
\end{equation}
Due to its simplicity and utility as a physical example, we begin with this fluid when we study the double copy prescription for the Type N fluid dual metrics in section \ref{PotFlowNis0}.

Other potential flows can also be written compactly in terms of $z$ and $\bar{z}$, using the form \eqref{AlphaPot}, as in Table \ref{tableTypeNflows}.  We will study the double copy of type N metrics dual to the generic potential flow fluid with potential \eqref{AlphaPot} in section \ref{generaltypeN} below.
\begin{table}[H]
\centering 
\begin{tabular}{c|c|c|c}
      \small{Type of fluid solution}&  Fluid Potential $\phi(t,x,y)$&$\phi({z,\bar{z}})$&$\Psi_4$ \\ \hline
      \small{Source/Sink} &$\alpha\,\, \textrm{ln}(x^2+y^2)$ &  $\alpha\ln (z\bar{z})$ & 
      2 $\alpha \,r^{-1}\bar{z}^{-2}$
      \\ \hline
      \small{Source to Sink (dipole)}&$\frac{\alpha \delta x}{x^2+y^2}$& $ \frac{\alpha \delta}{2}\frac{z+\bar{z}}{z\bar{z}}$ & $2\, r^{-1}\alpha\delta\bar{z}^{-3}$\\  \hline
            \small{Line Vortex}&$\alpha \arctan(y/x)$ &
           $\frac{\alpha}{2i}\ln\left(\frac{z}{\bar{z}}\right)$
             & $i \alpha\, r^{-1}\bar{z}^{-2}$\\\hline
   \small{Extensional flow}&$-\frac{\alpha}{2}(x^2-y^2)$ &$ -\frac{\alpha}{4}(z^2 +\bar{z}^2)$ & $ \frac{\alpha}{  r}$\\          
\end{tabular}   
\caption{Some examples of standard fluid solutions and the corresponding non-vanishing scalar $\Psi_4$ for type N solutions. For the dipole flow, $\delta$ refers to the distance between the source and the sink.}\label{tableTypeNflows}
\end{table}


\section{Type D Double Copy}\label{type-D}

\subsection{Weyl Double Copy }\label{TypeDWeylDCsection}
Now that we've obtained velocity and pressure fields that correspond to either Petrov type D or type N, we look to build the Weyl double copy (\ref{WeylDCrelation}) corresponding to the particular fluid solutions. Accordingly, we use our results for the Weyl scalars \eqref{epsilonWeylscalars} and the expansion of the Weyl spinor $C_{ABCD}$, given by (\ref{expandCABCD}). As we showed in section \ref{typeDfluidsSection}, the type D constraint leaves us with constant vorticity fluid solutions.  The time-independent solution (\ref{typeDvelocity}) and (\ref{typeDpressure}) takes the form
\begin{equation}\label{DvsandP}
v_x=-\omega y, \ \ \ \ \ \ \ v_y=\omega x, \ \ \ \ \ \ \ \ P=\omega^2\frac{(x^2+y^2)}{2}.
\end{equation}
From the expression for the Weyl scalars $\Psi_I$ for arbitrary velocity fields (\ref{epsilonWeylscalars}), we find that the solution (\ref{DvsandP}) leaves us with 
\begin{equation}\label{TypeDWeyl}
\Psi_2  = -i\epsilon^2\frac{\omega}{2r_c}+\calO(\epsilon^3),
\end{equation}
while all other $\Psi_I$ vanish to $\calO(\epsilon^3)$. Consequently, the Weyl spinor is $C_{ABCD}=6\Psi_2 o_{(A}o_B\iota_C\iota_{D)}$. 

Using the Weyl double copy as defined in (\ref{WeylDCrelation}), we find the zeroth copy scalar and single copy gauge field are, to lowest order in $\epsilon$, 
\begin{equation}\label{scalarAndfABforD}
S=\frac{i\omega r_c}{3}e^{2i\theta}, \ \ \ \ \ \ \ f_{AB}=e^{i\theta}\omega\begin{pmatrix}1&0\\0&-1\end{pmatrix},
\end{equation}
where $\theta$ is a constant (global) phase to be interpreted shortly. Since the double copy relation \eqref{WeylDCrelation} and the vanishing of all $\Psi_{I\neq 2}$ force $f_{AB}\propto o_A\iota_B$, the matrix structure of $f_{AB}$ here arises from the form of $o_A$ and $\iota_B$ as in (\ref{omicronsANDiotas}).

Next, we use the relation between the spacetime formalism and the spinor formalism as reviewed in appendix \ref{spinorappendix} to obtain the tensor form of the field strength $F^{\mu\nu}$ from the spinor $f_{AB}$.  These relationships necessitate a vierbein for the background on which the gauge fields live. We choose to interpret the gauge fields as living on the Rindler background
\begin{equation}\label{RindlerBackground}
ds_{(0)}^2=-rd\tau^2+2drd\tau+dx^2+dy^2,
\end{equation}
where the scalar satisfies the wave equation, $\nabla^{(0)\mu}\nabla^{(0)}_\mu S=\Box^{(0)}S=0.$ The $\nabla^{(0)}_\mu$ are the covariant derivatives with respect to (\ref{RindlerBackground}). From (\ref{tetradtovierbien}), we obtain the vierbeins

\begin{align}\label{flatEs2}
e_\mu^{(0),0}&=(-\sqrt{r},\frac{1}{\sqrt{r}},0,0), \nonumber\\
e_\mu^{(0),1}&=(0,-\frac{1}{\sqrt{r}},0,0), \\
e_\mu^{(0),2}&=(0,0,1,0), \nonumber\\
e_\mu^{(0),3}&=(0,0,0,1). \nonumber
\end{align}
Using (\ref{justF}) to obtain $F^{\mu\nu}$ in terms of $f_{AB}$, the Pauli matrices, and the vierbeins, we find the only nonzero components are
\begin{equation}\label{typeDFs}
    F^{\tau r}=-\omega\cos\theta, \ \ \ \ \ \ F^{xy}=-\omega\sin\theta.
\end{equation}

Recalling that the gauge field is in the $U(1)$ sector of Yang-Mills, the Maxwell equations 
\begin{equation}\label{typeDmax}
    \nabla^{(0)}_{\nu}F^{\mu\nu}=0, \ \ \ \ \ \ \ \nabla^{(0)}_{[\mu}F_{\rho\sigma]}=0,
\end{equation}
indeed show that the field strength (\ref{typeDFs}) is a vacuum solution. This is to be expected, since the fluid solutions are obtained by demanding the Einstein equations are satisfied in vacuum, $G_{\mu\nu}=0,$ so we expect the single copy to follow suit. In the classical double copy, it is possible for the spacetime to have a singularity that maps to a gauge field source, as the point mass maps to a point charge in the Schwarzschild solution \cite{Monteiro:2014cda} when parameters are chosen to turn off the dilaton \cite{Luna:2020adi,Kim:2019jwm}. Because Rindler space is free from singularities, no sources will be found on the gauge theory side, consistent with (\ref{typeDmax}).

\subsection{Effective electric and magnetic fields}\label{EBforTypeD}

Interpreting the single copy gauge field strength (\ref{typeDFs}) as a Maxwell solution allows us to discuss the electric and magnetic fields whose double copy generates the metric dual to a constant vorticity fluid.%
\footnote{Note that unlike references \cite{Ilderton:2018lsf} and \cite{Andrzejewski:2019hub}, which discuss gauge and gravity solutions with vorticity, we are discussing metrics dual to fluids with vorticity.}
 These fields are defined covariantly by 
\begin{equation}\label{covariantEandB}
E_\nu= F_{\nu\mu}\xi^\mu, \ \ \ \ \ \ \ \ B_\nu=\frac{1}{2}\varepsilon_{\mu\nu\rho\sigma}F^{\rho\sigma}\xi^\mu,
\end{equation}
where $\xi$ is the (timelike) Killing vector $\xi=\partial_\tau$. For the field strength under consideration, we find
\begin{equation}\label{EBfromEpsilonTypeD}
E_\nu = \omega\cos\theta\delta_\nu^r, \ \ \ \ \ \ \ \ B_\nu = -\omega\sin\theta\delta_\nu^r.
\end{equation}

We interpret these fields by choosing the global phase to be $\theta=\frac{3\pi}{2}$, which leaves us with a constant magnetic field pointing in the $r$ direction, perpendicular to the $x-y$ plane. Under this choice of $\theta$, the classical vector potential $\vec{A}$, which constructs the magnetic field by $\vec{B}=\nabla\times\vec{A}$, coincides with the velocity fields directly: $\vec{A}\propto\vec{v}$.  Since the magnetic field is unchanged when the vector potential shifts by a constant, we see that the single copy gauge fields will similarly be unchanged when we shift the velocity by a constant.

We also compute the electromagnetic stress tensor
\begin{equation}\label{emStressEqn}
T^{\rho\sigma}=F^\rho_{\phantom{\mu}\mu}F^{\sigma\mu}-\frac{1}{4}g^{\rho\sigma}F_{\mu\nu}F^{\mu\nu},
\end{equation}
finding the nonzero components
\begin{equation}\label{TypeDTmunu}
T^{\tau r} = -\frac{\omega^2}{2},\quad T^{rr}=-\frac{r\omega^2}{2},\quad T^{xy}=\frac{\omega^2}{2}.
\end{equation}
The associated energy with respect to the Killing vector $\xi$ is given by
\begin{equation}\label{TypeDT00}
T^{\mu \nu}\xi_\mu\xi_\nu=\omega^2r/2,
\end{equation}
while the spatial components of the Poynting vector, from $T^{\mu \nu}\xi_\mu$, become zero.

 Physically, we can understand the fluid (\ref{DvsandP}) as the solution inside of a slowly rotating cylinder with its axis along the $r$-direction and no-slip boundary conditions at the wall, where we have taken the radius of the cylinder to be large (with respect to all other scales in the problem). The corresponding single copy gauge field, $\vec{B}=\omega \hat{r}$, matches the uniform magnetic field along the axis of a solenoid with $n$ turns per unit length whose current $I$ is proportional to $\omega/n$. The axis of the solenoid is aligned with the axis of the cylinder containing the fluid.%
\footnote{The velocity fields rotate counter-clockwise in the $x-y$ plane. After exchanging the vorticity parameter with a current parameter, the resulting magnetic field then points along positive $\hat{r}$, consistent with choosing $\theta=3\pi/2.$} %
The double copy mapping therefore associates the vorticity of the fluid with the magnitude of the current sourcing the magnetic field.   The field moreover has energy dependent on the radial location $r$, but has vanishing Poynting vector as expected for a pure magnetic field. In addition, we see from (\ref{scalarAndfABforD}) that the zeroth copy $S$ plays a passive role in that it trivially solves the wave equation. We thus find that all of the nontrivial information that is mapped through the double copy is contained in the field strengths $f_{AB}$ or $F^{\mu\nu}$ for the type D spacetime.

\subsection{Weyl Double Copy in the Near Horizon Expansion}
The hydrodynamic limit can be related to a near horizon expansion of the metric by rescaling the metric as in  \eqref{rescale} \citep{Bredberg:2011jq}. Since the full fluid solution \eqref{DvsandP} does not actually lie in the hydrodynamic regime%
\footnote{
The fluid solution \eqref{DvsandP} is only in the hydrodynamic regime \eqref{hydrolimit} for $x,\, y \sim\epsilon^{-1}$ while the vorticity satisfies $\omega\sim\epsilon^2$. For either small $x,\,y$ or large vorticity, the solution exits the hydrodynamic regime, although of course it still solves Navier-Stokes.  Because of this technicality, the metric \eqref{mrc} is not trustable for small $x,\, y$.  However, in the near-horizon expansion, because of the rescaling \eqref{rescale}, the fluid solution does not need to be in the hydrodynamic regime, since this expansion is rewritten explicitly in terms of the hatted coordinates in \eqref{hattedcoords} that are of $O(1)$. Here we explore an explicit realization of the near-horizon expansion, for completeness, as provided in equation \eqref{lambdaexpansion}.}%
, we repeat here the same analysis as in section \ref{TypeDWeylDCsection}, repeated in the near horizon expansion \eqref{lambdaexpansion}.  We again find the same results.

Using the tetrad \eqref{lambdatetrad}, we find the Weyl scalars for the near horizon metric \eqref{lambdaexpansion} with the constant vorticity fluid \eqref{DvsandP}.  The only nonzero Weyl scalar is
\begin{equation}
\Psi_2 = \frac{i\omega}{2} + \calO(\lambda).
\end{equation}
All other Weyl scalars vanish at $\calO(1)$, and have contributions from neglected pieces of the metric at $\calO(\lambda)$ or higher.  Following the method in section \ref{TypeDWeylDCsection}, we identify the zeroth copy scalar and single copy gauge field spinor:
\begin{equation}\label{fABlambda}
    S=\frac{1}{3}e^{i(\pi+2\theta)}, \ \ \ \ \ \ \ f_{AB}=\omega e^{i\theta}\begin{pmatrix}1&0 \\ 0&-1\end{pmatrix}.
\end{equation}
As before, we obtain the appropriate flat space vierbien by setting the velocities and pressures to zero in the full tetrad and using eq. \ref{tetradtovierbien}; we find
\begin{equation}\label{flatEs3}
    e_{ \ \ \mu}^{(0), a}=\begin{pmatrix}\frac{r+\lambda}{2\lambda}&\frac{r-\lambda}{2\lambda}&0&0 \\
    -1&-1&0&0 \\
    0&0&1&0\\
    0&0&0&1
    \end{pmatrix}.
\end{equation}
Using this flat space vierbien the gauge field strength tensor in the $\lambda$ expansion becomes 
\begin{equation}
    F^{\tau r}=-\omega\cos\theta, \ \ \ \ \ \ F^{xy}=-\omega\sin\theta,
\end{equation}
which should be thought of as living on a flat Rindler background.  We then identify effective electric and magnetic fields, which are identical to the previous result  \eqref{EBfromEpsilonTypeD} obtained in the hydrodynamic limit:
\begin{equation}
    E_{\nu}=\omega\cos\theta\,\, \delta_{\nu}^{\,\,r}, \ \ \ \ \ \ B_{\nu}=-\omega\sin\theta\,\,\delta_{\nu}^{\,\,r}.
\end{equation}
%

\section{Type N Weyl Double Copy}\label{type-N}
In this section we will analyze the single copy gauge fields and zeroth copy scalar fields corresponding to the metrics dual to potential flow fluids. As we saw in section (\ref{typeNfluids}), these potential flows are the most general solution whose dual metrics satisfy the Petrov type N constraint.  As potential flows, their velocity can be written as the gradient of a scalar potential, $v_i=\partial_i\phi$, where $\phi$ satisfies Laplace's equation in $\mathbb{R}^2$. For convenience, we defined $z=x+iy$ and its conjugate $\bar{z}$ so that we may write the Laplacian as $\partial^2=\partial_z\partial_{\bz}$, decomposing the scalar potential as $\phi(z,\bz)=f(z)+\bar{f}(\bar{z})$. The resulting Weyl scalar, $\Psi_4$, is given by (\ref{psi4}), and all others vanish. Therefore the Weyl double copy should satisfy 
\begin{equation}\label{typeNdc}
C_{ABCD}=-\frac{2}{r	}\partial^2_{\bz}\bef(\bz)o_A o_B o_C o_D = \frac{1}{S}f_{AB}f_{CD}.
\end{equation}

\subsection{Planar Extensional Flows}\label{PotFlowNis0}
Let us start with the simple case of planar extensional flow, where $\phi(z,\bar{z})=-\frac{\alpha}{4}(z^2+\bar{z}^2)$ with $\alpha$ a real constant. The corresponding velocity fields are (\ref{TypeNfluidSolutions}) $v_x=-\alpha x$ and $v_y=\alpha y$. 

We can satisfy the double copy relation (\ref{typeNdc}) by choosing
\begin{equation}\label{typeNfandS}
S=\frac{e^{2i\theta}}{\alpha}, \qquad f_{AB}=\frac{ e^{i\theta}}{\sqrt{r}}\begin{pmatrix}
1&1\\1&1
\end{pmatrix},
\end{equation}
where we again allow for a global phase $\theta.$ Here, since we have $\Psi_{I\neq 4}=0$, we have $f_{AB}\propto o_A o_B$, therefore the matrix structure in (\ref{typeNfandS}) arises from (\ref{omicronsANDiotas}). Although we could make another choice for $S$, this constant choice trivially satisfies $\Box^{(0)}S=0$, and $f_{AB}$ is the only choice which will satisfy the gauge field equations as we show below.

As for the type D case, we specify our background spacetime by using \eqref{tetradtovierbien} to find the vierbeins corresponding to the tetrads used to compute $\Psi_4$, and then setting $v_i=P=0$.  The resulting vierbeins turn out to have the same form as (\ref{flatEs2}). We then obtain the gauge field strength tensor via (\ref{justF}), finding 
\begin{equation}\label{TypeNGaugeField}
F^{r x}=-\sin\theta, \qquad F^{r y}=-\cos\theta, \qquad F^{\tau x}=-\frac{2\sin\theta}{r}.
\end{equation}
As in the type D case, since this field strength has no nontrivial color factor dependence, we treat it as an effective Maxwell field; indeed it satisfies the vacuum Maxwell equations over the Rindler background (\ref{RindlerBackground}) for arbitrary $\theta$. 

We obtain the electric and magnetic fields using the covariant expressions (\ref{covariantEandB}), yielding 
\begin{equation}\label{typeNE}
E_\nu =(0,0,\sin\theta,-\cos\theta)
\end{equation}
and
\begin{equation}\label{typeNB}
B_\nu = (0,0,\cos\theta,-\sin\theta).
\end{equation}
Again, as in the type D case, we choose $\theta=3\pi/2$ as a convenient parametrization; picking another $\theta$ will just result in a rotation in the $x,\, y$ plane.  Computing the electromagnetic stress tensor \eqref{emStressEqn}, we find
\begin{equation}
T^{\tau\tau}=\frac{4}{r^2}, \quad T^{\tau r}=\frac{2}{r}, \quad T^{rr}=1.
\end{equation}
The energy becomes
\begin{equation}
T^{\mu\nu}\xi_\mu\xi_\nu=1,
\end{equation}
while the spatial components of the Poynting vector become
\begin{equation}\label{typeNPoynting}
S^i = -\delta^i_r. 
\end{equation}
We interpret this gauge field as the single copy field necessary to build up any fluid which has a potential component.  Since any two-dimensional vector field can be decomposed, via the two-dimensional version of Helmholtz decomposition, we can write the velocity field as
\begin{equation}\label{2dHelm}
v_i = \partial_i \phi + \epsilon_{ijk} \partial_j A_k,
\end{equation}
where the vector fluid potential for the two-dimensional case satisfies $\vec{A}=|A|(\hat{x}\times\hat{y})$, and $i,j,k$ run over the directions $x$ and $y$ as well as the direction $\hat{x}\times\hat{y}$.
For the potential flows whose gravity duals are type N, we have only the first term; that is, $|A|=0$.  Most of the information in $\phi$ will be carried instead by the scalar $S$, so the field profile \eqref{TypeNGaugeField} is only building up the fluid-dual spacetime necessary to support a velocity field with a nonzero $\partial_i \phi$ term.

The nonzero Poynting vector \eqref{typeNPoynting} indicates the dissipative nature of these flows.  The gravitational dual is carrying energy away from the $r=r_c$ hypersurface, towards the null horizon, satisfying the infalling Rindler boundary conditions that underlie the derivation of the fluid-dual metric \eqref{mrc}.  The same flow of energy towards the null horizon arises in the Poynting vector aligned in the $-\hat{r}$ direction.

\subsection{General potential flows}\label{generaltypeN}
As we will show, the analysis in section \ref{PotFlowNis0} will work very similarly for a potential flow $\phi=f(z)+\bar f(\bar z)$ with generic holomorphic function $f(z)$.

Since $\Box^{(0)}$ on the Rindler background (\ref{RindlerBackground}) will give zero when acting on any function which is a sum of holomorphic and antiholomorphic terms independent of $\tau$ and $r$, we can satisfy the type N Weyl double copy relation (\ref{typeNdc}) for the metric dual to a generic potential flow with 
\begin{equation}\label{TypeNGeneralCopy}
S=-\frac{e^{2i\theta}}{2\partial^2_{\bz}\bef(\bz)}, \qquad f_{AB}=\frac{ e^{i\theta}}{\sqrt{r}}\begin{pmatrix}
1&1\\1&1
\end{pmatrix}.
\end{equation}
It is now the case that $\Box^{(0)}S=0$ is nontrivially satisfied, and the resulting gauge field strength is unchanged from the analysis for the planar extensional flow. Thus for all potential flow fluids, such as those in Table \ref{tableTypeNflows}, the Weyl double copy admits the same single copy gauge field as in the extensional flow, \eqref{TypeNGaugeField}.  The information for a potential flow on the fluid side resides entirely in the potential $\phi$; similarly, under the double copy prescription, we find that the information from the potential resides entirely in the zeroth copy scalar field $S$, whereas the single copy gauge field is the same for all potential flows.

Since the single copy field profile is again \eqref{TypeNGaugeField}, our interpretation of this field as building the fluid-dual spacetime for fluids with nonzero potential terms holds again.  We do note that the fields \eqref{typeNE} and \eqref{typeNB} are constant; we expect that inclusion of higher order terms in the $\epsilon$ expansion could alter this result, since here we are really considering only a hydrodynamic expansion in small $\epsilon$ around the original $r=r_c$ cutoff surface.

\section{Discussion}
\label{discussions}
We have used the Weyl double copy prescription to find the single copy gauge fields and zeroth copy scalar fields arising from two classes of fluid-dual metrics.  The first class, fluids with constant vorticity, maps to spacetime metrics with Petrov type D. The second class, potential flow fluids, maps to spacetime metrics with Petrov type N.  For the type D spacetimes dual to fluids with constant vorticity, we find an (effectively abelian) dual gauge field with vanishing Poynting vector. For the type N spacetimes dual to potential flows, we find a gauge field whose Poynting vector points in towards the Rindler horizon, indicating that the dissipation in these fluids maps in the spacetime to energy flowing across the horizon due to the infalling boundary conditions there.

We also saw that the single and zeroth copy fields mapping to the two sets of fluid-dual metric classes store their information differently. In the type D case, the vector potential for the magnetic field corresponds to the fluid velocity profile, while the zeroth-copy scalar field is just a constant; only the single-copy gauge field is carrying nontrivial information about the fluid.  For type N spacetimes, the story is in some sense opposite: the nontrivial components of the fluid are entirely due to the potential, which shows up only in the zeroth-copy scalar field.  Here, the gauge field is fixed and appears to be the field necessary to build the fluid-dual spacetime for all potential flow fluids.

In fact, the two fluid classes we have studied fall into two simple classes under the Helmholtz decomposition, which rewrites the fluid vector field in terms of its rotational component and its irrotational or potential component, as in \eqref{2dHelm}.  The constant vorticity solutions which map to type D spacetimes have $\phi=0$ while the potential flow solutions that map to type N spacetimes have $\vec{A}=0$.  Under the double copy prescription, solutions with nonzero $\vec{A}$ map to a nontrivial gauge field whose behavior depends on the fluid velocity, but to a constant (trivial) zeroth copy scalar.  Similarly, solutions with nonzero $\phi$ all map to the same gauge field \eqref{TypeNGaugeField}, so instead the zeroth copy scalar carries the fluid information: it is proportional to the second derivative of the fluid potential as in \eqref{TypeNGeneralCopy}. Consequently, we propose that any fluid-dual metric may be mapped to a single copy gauge field and zeroth copy scalar, each of which is a sum of the corresponding pieces from the rotational and irrotational components in the Helmholtz decomposition.  We hope to explore this idea in future work.

We should note throughout that we work only to the lowest order in a perturbative expansion (mainly the hydrodynamic expansion).  A more complete treatment may require understanding of the double copy prescription beyond a linear order; all double Kerr-Schild prescriptions are essentially linear due to the linearization of the equations of motion in those coordinates.  The Weyl double copy itself is not linear in nature, but is unclear how it might relate to more advanced treatments that would go beyond a perturbative expansion as in \cite{Luna:2016hge}, such as the convolution prescription in \cite{Luna:2020adi,Kim:2019jwm}.  Further development of this convolution prescription to include algebraically special spacetimes would be of interest.

The double-copy treatment in the fluid-gravity duality context may also be amenable to analysis using solution generating techniques.  For example, the Ehler's transformations as implemented in \cite{Berkeley:2012kz} for fluids and further studied in \cite{Alawadhi:2019urr,Huang:2019cja} in the context of the double copy could allow access to a larger set of double-copy treatments for fluid-dual spacetimes. Indeed, such an analysis could shed light on the nature of single and zeroth copies for such spacetimes.

Since fluid-gravity duality itself can be understood from an AdS-CFT perspective (including the cutoff-prescription formulation, whose relationship to AdS-CFT was first understood in \cite{Brattan:2011my}), we hope the mapping here from fluid solutions to gravities and then through the double copy prescription to gauge theories (and scalars) can provide perspective both regarding the relationship of the double copy prescription to AdS-CFT duality, and also the understanding of fluid-gravity duality itself, including a deeper understanding of fluids as in \cite{Haehl:2015foa}.

\subsection*{Acknowledgements}
We would like to thank J.J. Carrasco, Damien Easson, and Andres Luna for useful and enlightening conversations while this work was in progress. This work is supported by the U.S. Department of Energy under grant number DE-SC0019470.

\appendix
\section{Spinor formalism}\label{spinorappendix}

In our notation, spacetime indices are given by $\{\mu,\nu,\gamma,...\},$ frame indices by $\{a,b,c,...\}$ and the spinor indices as $\{A,B,C,...\}$ with their conjugates $\{\dA,\dB,\dC,...\}$. The essential objects that translate between the spinor and tensor formalisms are the Pauli 4-vectors

\begin{equation}\label{framesigmas}
\sigma^a_{A\dA} = \frac{1}{\sqrt{2}}\big(1,\vec{\sigma}\big)_{A\dA}, \ \ \ \ \ \ \vec{\sigma}=(\sigma_x,\sigma_y,\sigma_z).
\end{equation}
The $\vec{\s}$ are the standard $SU(2)$ generators,
\begin{equation}
\s_x = \begin{pmatrix}
0&1 \\ 1&0
\end{pmatrix}, \ \ \ \ \ \ \s_y=\begin{pmatrix}
0 & -i \\ i &0
\end{pmatrix}, \ \ \ \ \ \ \s_z = \begin{pmatrix}
1&0 \\ 0&-1
\end{pmatrix}.
\end{equation}

A spacetime vector is obtained from a frame vector by $V_\mu=e_\mu^{ \ a}V_a,$ where the $e_\mu^{ \ a}$ are vierbeins that construct the full metric as $g_{\mu\nu}=e_\mu^{ \ a}e_\nu^{ \ b}\eta_{ab}$. Here, $\eta_{ab}=\eta^{ab}=\text{diag}(\text{-}1,1,1,1)$. The frame indices are raised and lowered with the diagonal Minkowski space $\eta_{ab}$, while spinor indices are raised and lowered with a Levi-Civita symbol, which we define as 
\begin{equation} 
\varepsilon^{AB}=-\varepsilon_{AB}=\begin{pmatrix}
0&1\\\text{-}1&0
\end{pmatrix}.
\end{equation}
A vector can be written in spinor indices or in frame indices using (\ref{framesigmas});
\begin{equation} \label{SpinorToVector}
V_{A\dA}=V_a\sigma^a_{A\dA}, \qquad \Leftrightarrow \qquad V_a = \sigma_{aA\dA}V^{A\dA},
\end{equation}
where $\sigma_{aA\dA}=\eta_{ab}\s^b_{A\dA}$ and $V^{A\dA}=\varepsilon^{AB}V_{B\dB}\varepsilon^{\dB\dA}$.
The (inverse) vierbein constructs the Pauli 4-vector in spacetime indices $\sigma^\mu_{A\dA}=e^{\mu}_{ \ a}\sigma^a_{A\dA}$ which, with its inverse $\sigma_\mu^{A\dA}=g_{\mu\nu}\varepsilon^{AB}\sigma^\nu_{B\dB}\varepsilon^{\dB\dA}$, satisfies
\begin{equation}
\sigma^{\mu}_{A\dA}\sigma_\nu^{A\dA}=\delta^\mu_\nu, \ \ \ \ \ \ \sigma^\mu_{A\dA}\sigma_\mu^{B\dB}=\delta^B_A \delta^{\dB}_{\dA}.
\end{equation}
Any tensor can be written as its spinor counterpart using the index doubling procedure. The Weyl tensor $W_{\mu\nu\lambda\gamma}$ becomes
\begin{equation}
W_{\mu\nu\lambda\gamma}\rightarrow \ W_{A\dA B\dB C\dC D\dD}=C_{ABCD}\varepsilon_{\dA\dB}\varepsilon_{\dC\dD}+\bar{C}_{\dA \dB\dC\dD}\varepsilon_{AB}\varepsilon_{CD},
\end{equation}
where the $C_{ABCD}$ and $\bar{C}_{\dA\dB\dC\dD}$ are symmetric in their indices and related by complex conjugation. The object\footnote{Brackets denote antisymmetrization, and we use the convention $[A,B]=AB-BA$.}
\begin{equation}\label{bigsigmas}
\sigma^{\mu\nu}_{AB} = \sigma^{[\mu}_{A\dC}\bar{\sigma}^{\nu] \ \dC C}\varepsilon_{CB}, \ \ \ \ \ \text{with} \ \ \ \ \  \bar{\sigma}^{\mu A\dA} =e^{\mu}_{ \ a}\bar{\sigma}^{a\dA A}, \ \ \ \ \ \ \bar{\sigma}^{a\dA A}= \frac{1}{\sqrt{2}}\big(1,-\vec{\sigma}\big)^{\dA A}
\end{equation}
serves to directly obtain the spinor form of a given tensor. For the Weyl spinor, 
\begin{equation}\label{WeylSpinorAndTensor}
C_{ABCD}=\frac{1}{4}W_{\mu\nu\lambda\gamma}\sigma^{\mu\nu}_{AB}\sigma^{\lambda\gamma}_{CD}.
\end{equation}
For the field strength tensor $F_{\mu\nu},$ we write
\begin{equation}
F_{\mu\nu}\rightarrow \ F_{A\dA B\dB}=f_{AB}\varepsilon_{\dA\dB}+\bar{f}_{\dA\dB}\varepsilon_{AB},
\end{equation}
where the spinor field strength can be computed as
\begin{equation}\label{fAB}
f_{AB}=\frac{1}{2}F_{\mu\nu}\sigma^{(0)\mu\nu}_{AB},
\end{equation}
which is also symmetric in its spinor indices. In the above expression, the zero superscript is meant to remind that since $F_{\mu\nu}$ lives on flat space, the vierbein that's used to construct the $\s^{\mu}_{A\dA}$ in (\ref{fAB}) is that which constructs the flat space,
\begin{equation}
\s^{(0)\mu}_{A\dA}=e^{(0)\mu}_{ \ \ a}\s^a_{A\dA}, \ \ \ \ \ \ \ e_\mu^{(0)a}e_\nu^{(0)b}\eta_{ab}=g^{(0)}_{\mu\nu}.
\end{equation}
For example in section \ref{TypeDWeylDCsection}, $g^{(0)}_{\mu\nu}$ is Rindler space (\ref{RindlerBackground}) and the $e^{(0)a}_{\mu}$ are (\ref{flatEs2}). The vierbeins that are used to build $\s^{\mu\nu}_{AB}$ in (\ref{WeylSpinorAndTensor}) instead construct the full spacetime. For conciseness we will drop the 0-superscript in what follows.

To invert (\ref{fAB}), it is tedious though straightforward to show
\begin{equation}\label{Ff}
F^{\mu\nu}-\frac{i}{2}\frac{\varepsilon^{\mu\nu\alpha\beta}}{\sqrt{-g}}F_{\alpha\beta}=\s^{\mu A\dD}f_{AB}\varepsilon^{BD}\bar{\s}^\nu_{\dD D},
\end{equation}
where $g=\text{det}g_{\mu\nu}$. $F^{\mu\nu}$ can be obtained directly by adding the complex conjugate of the right hand side in (\ref{Ff}), yielding
\begin{equation}\label{justF}
F^{\mu\nu}=\frac{1}{2}\Bigg[\s^{\mu A\dD}f_{AB}\varepsilon^{BD}\bar{\s}^\nu_{\dD D}+\s^{*\mu\dA D}\bar{f}_{\dA\dB}\varepsilon^{\dB\dD}\bar{\s}^{*\nu}_{D\dD}\Bigg].
\end{equation}
For the second term in (\ref{justF}), the $\sigma^*$ denotes standard complex conjugation, i.e. 
\begin{equation} 
\s^{*a}_{\dA A}=\frac{1}{\sqrt{2}}\Big(1,\s_x ,-\s_y, \s_z\Big)_{\dA A}.
\end{equation}

\section{Newman-Penrose formalism}\label{NPappendix}
We now briefly describe the Newman-Penrose (NP) formalism which we use to compute geometric quantities of interest such as the Weyl spinor. The NP formalism utilizes spinor language in order to simplify computations (\cite{Penrose:1987uia},\cite{Penrose:1986ca},\cite{Stephani:2003tm}). There primarily are four sets of objects of interest for us in the NP formalism. Briefly, one rewrites the metric in terms of a tetrad set, this tetrad set then is used to compute spin coefficients\footnote{We utilize the method outlined in \cite{Cocke1989} to obtain spin-coefficients, this approach comes with the computational benefit of replacing certain covariant derivatives with partial derivatives)},
\begin{equation}
\label{tetradmetric}
 g_{\mu\nu} = -l_{(\mu}n_{\nu)} + m_{(\mu}\overline{m}_{\nu)}.
\end{equation}
Bilinears of the spin coefficients then give the set of Weyl scalars $\{\Psi_0,\Psi_1,\Psi_2,\Psi_3,\Psi_4,\}$,
\begin{equation}\label{NP}
    \begin{split}
        \Psi_{0}&=D\sigma-\delta\kappa-(\rho+\bar{\rho}+3\varepsilon+\bar{\varepsilon})\sigma+(\tau-\bar{\pi}+\bar{\alpha}+3\beta)\kappa \\
        \Psi_{1}&=D\beta-\delta\varepsilon-(\alpha+\pi)\sigma-(\bar{\rho}-\bar{\varepsilon})\beta+(\mu+\gamma)\kappa+(\bar{\alpha}-\bar{\pi})\varepsilon \\
        \Psi_{2}&=D\mu-\delta\pi+(\varepsilon+\bar{\varepsilon}-\bar{\rho})\mu+(\bar{\alpha}-\beta-\bar{\pi})\pi+\nu\kappa-\sigma\lambda-R/12 \\
        \Psi_{3}&=\bar{\delta}\gamma-\Delta\alpha+(\rho+\varepsilon)\nu-(\tau+\beta)\lambda+(\bar{\gamma}-\bar{\mu})\alpha+(\bar{\beta}-\bar{\tau})\gamma \\
        \Psi_{4}&=\bar{\delta}\nu-\Delta\lambda -(\mu+\bar{\mu}+3\gamma-\bar{\gamma})\lambda+(3\alpha+\bar{\beta}+\pi-\bar{\tau})\nu,
    \end{split}
\end{equation}
where the following are directional derivatives, 
\begin{equation}
    D=l^{\mu}\nabla_{\mu}, \ \ \ \ \Delta=n^{\mu}\nabla_{\mu}, \ \ \ \ \delta=m^{\mu}\nabla_{\mu}, \ \ \ \ \bar{\delta}=\bar{m}^{\mu}\nabla_{\mu}.
\end{equation}
Finally in terms of these Weyl scalars one can rewrite the Weyl Spinor.
\begin{equation}
    C_{ABCD}=\Psi_0 \iota_{A}\iota_B\iota_C\iota_D-4\Psi_1 o_{(A}\iota_B\iota_C\iota_{D)}+6\Psi_2 o_{(A}o_B\iota_C\iota_{D)}-4\Psi_3 o_{(A}o_Bo_C\iota_{D)}+\Psi_4 o_Ao_Bo_Co_D\\
\end{equation}
Finally in order to test the algebraic speciality of the spacetime we compute tetrad invariant combinations of the Weyl scalars; the equation below is equivalent to \eqref{Weylinvariants} in the main text as included here for completeness:
\begin{equation}\label{AppendInvar}
    \begin{split}
        I&\equiv\Psi_0\Psi_4-4\Psi_{1}\Psi_{3}+3\Psi_{2}^{2}, \\
        J&\equiv \begin{vmatrix} 
        \Psi_{4}&\Psi_{3}&\Psi_{2}\\
        \Psi_{3}&\Psi_2 &\Psi_1 \\
        \Psi_{2}&\Psi_{1} &\Psi_{0}
        \end{vmatrix} ,\\
        K&\equiv \Psi_{1}\Psi_{4}^{2}-3\Psi_{4}\Psi_{3}\Psi_{2}+2\Psi_{3}^{3}, \\
        L&\equiv \Psi_{2}\Psi_{4}-\Psi_{3}^{2}, \\
        N&\equiv 12L^{2}-\Psi_{4}^{2}I.
    \end{split}
\end{equation}
The spinors ${o_A,\iota_A}$ are related to the frame metric choice one makes. We will make explicit this connection now. The metric written in terms of vierbiens has the form,
\begin{equation}
    g_{\mu\nu} = e_{\mu}^{\,\,a} e_{\nu}^{\,\,b} \eta_{ab} \,\,\,\, \text{where}\,\,\,\,\eta_{ab} = \text{diag} \{-1,1,1,1\}.
\end{equation}
The frame metric $\eta_{ab}$ can itself be written as outer products of a tetrad set, this will allow us to make identifications between the vierbiens and the tetrad set.
\begin{equation}
\begin{split}
    \eta_{ab} &= -\hat{l}_{(a}\hat{n}_{b)} + \hat{m}_{(a}\hat{\overline{m}}_{b)}\\
    \implies g_{\mu\nu} &=  e_{\mu}^{\,\,a} e_{\nu}^{\,\,b} (-\hat{l}_{(a}\hat{n}_{b)} + \hat{m}_{(a}\hat{\overline{m}}_{b)})\\
    \implies g_{\mu\nu} &= -l_{(\mu}n_{\nu)} + m_{(\mu}\overline{m}_{\nu)}
\end{split}    
\end{equation}
Where in the last step we have made the identifications, 
\begin{equation}
\label{vtot}
    e_{\mu}^{\,\,a}\hat{l}_a = l_{\mu} \hspace{0.3cm} e_{\mu}^{\,\,a}\hat{n}_a = n_{\mu} \hspace{0.3cm} e_{\mu}^{\,\,a}\hat{m}_a = m_{\mu} \hspace{0.3cm} e_{\mu}^{\,\,a}\hat{\overline{m}}_a = \overline{m}_{\mu} 
\end{equation}
Now the tetrad set that reproduces the Minkowski frame metric is, 
\begin{equation}
\begin{split}
    \hat{l}_a &= \frac{1}{\sqrt{2}}\{1,-1,0,0\}\\
    \hat{n}_a &= \frac{1}{\sqrt{2}}\{1,1,0,0\}\\
    \hat{m}_a &= \frac{1}{\sqrt{2}}\{0,0,i,1\}\\
    \hat{\overline{m}}_a &= \frac{1}{\sqrt{2}}\{0,0,-i,1\}\\    
\end{split}    
\end{equation}
The expression \ref{vtot} can be inverted to go from tetrads to vierbiens via the following, 

\begin{equation}\label{tetradtovierbien}
\begin{split}
&e_{\mu}^{\,\,0} = \frac{1}{\sqrt{2}}(l_{\mu}+n_{\mu}) \qquad \,\,\,\,\, e_{\mu}^{\,\,1} = \frac{1}{\sqrt{2}}(l_{\mu}-n_{\mu})\\
&e_{\mu}^{\,\,2} = \frac{i}{\sqrt{2} }  (\bar{m}_{\mu}-m_{\mu}) \qquad
e_{\mu}^{\,\,3} =  \frac{1}{\sqrt{2} }  (m_{\mu}+\bar{m}_{\mu})
\end{split}
\end{equation}

In order to obtain the spinors we write these four vectors in an SL(2,C) representation by contracting them with relevant $\sigma$ matrices. Note in our conventions we have $\sigma^a_{A\dot{A}}=\{\mathbb{I},\vec{\sigma}\}$, while the curved space equivalents can be obtained by contracting these with vierbiens (i.e. $ \sigma^{\mu}_{A\dot{A}} = e^{\mu}_{\,\,\,a}\,\sigma^a_{A\dot{A}}$).  For eg., for $\{o_A,\iota_A\}$ we have,
\begin{equation}\label{omicronsANDiotas}
\begin{split}
   o_A o_{\dot{A}} &\equiv \hat{l}_a\, \sigma ^a_{A\dot{A}} = \frac{1}{2}  \begin{pmatrix}
  1 & 1\\
  1 & 1 \\ \end{pmatrix}\\
  &\implies o_A = \frac{1}{\sqrt{2}}\{1,1\}\\
   \iota_A \iota_{\dot{A}} &\equiv \hat{n}_a\, \sigma ^a_{A\dot{A}} = \frac{1}{2}  \begin{pmatrix}
  1 & -1\\
  -1 & 1 \\ \end{pmatrix}\\
  &\implies \iota_A = \frac{1}{\sqrt{2}}\{1,-1\}\\  
\end{split}
\end{equation}
 
 Further noting that one can transform from SL(2,C) left to right by complex conjugation we use the convention, 
 \begin{equation}
     (o_A)^* \equiv o_{\dot{A}}
 \end{equation}
This further verifies that the two remaining contractions will hold the following relations correctly, 
\begin{equation}\label{outerOmicronIota}
\begin{split}
  \hat{m}_a\, \sigma ^a_{A\dot{A}} &= \frac{1}{2}  \begin{pmatrix}
  1 & 1\\
  -1 & -1 \\ \end{pmatrix} =    \iota_A o_{\dot{A}}\\
 \hat{\overline{m}}_a\, \sigma ^a_{A\dot{A}} &= \frac{1}{2}  \begin{pmatrix}
  1 & -1\\
  1 & -1 \\ \end{pmatrix} =   o_A \iota_{\dot{A}} \\
\end{split}
\end{equation}

\section{Tetrads in the hydrodynamic and the near horizon expansions}\label{TetradsAppendix}

In the hydrodynamic limit as discussed in section 2 in the body of the paper, the velocities and the pressure must satisfy the scaling \eqref{hydrolimit}, 
where $i$ runs over $x$ and $y$, the spacelike coordinates on the cutoff surface $r=r_c$. We can make this scaling of derivatives explicit by making the following identifications to simplify keeping track of the $\epsilon$ orders:
\begin{equation}
    v_i \to v_{i,\epsilon} \equiv v_i(\epsilon^2 \tau, \epsilon x_i), \qquad P \to P_{\epsilon} \equiv P(\epsilon^2 \tau, \epsilon x_i). 
\end{equation}
With these identifications having been established we can now write out the tetrad set we use for the computation in the hydrodynamic expansion:
\begin{equation}
\label{epsilontetrad}
\begin{aligned} 
    l_{\mu} &= \left\{-\frac{\sqrt{r}}{\sqrt{2}},0,0,0\right\}\\&\hspace{0.7cm} +\epsilon^2 \left\{-\frac{\sqrt{r} \left(4 r_c P_{\epsilon}+(3 r-2
   r_c) \left(v_{x,\epsilon}^2+v_{y,\epsilon}^2\right)\right)}{4 \sqrt{2} r_c^2},\frac{\sqrt{r}
   \left(v_{x,\epsilon}^2+v_{y,\epsilon}^2\right)}{2 \sqrt{2} r_c^2},0,0\right\}+\calO(\epsilon^3);
\\
    n_{\mu}&=\left\{-\sqrt{\frac{r}{2}},\sqrt{\frac{2}{r}},0,0\right\} +  \epsilon^2\left\{-\frac{(r-2 r_c) \left(4 r_c P_{\epsilon}+r 
   \left(v_{x,\epsilon}^2+v_{y,\epsilon}^2\right)\right)}{4 r_c^2 \sqrt{2 r}  },0,0,0\right\} + \calO(\epsilon^3);
\\
   m_{\mu}&=\left\{0,0,-\frac{i}{\sqrt{2}},\frac{1}{\sqrt{2}}\right\} 
   +\epsilon^2 \left\{0,0,\frac{i (r-r_c) v_{x,\epsilon}^2}{2
   \sqrt{2} r_c^2},\frac{i (r-r_c) v_{y,\epsilon} \left(2 v_{x,\epsilon}+i v_{y,\epsilon}\right)}{2 \sqrt{2} r_c^2}\right\}+\calO(\epsilon^3);
\\
   \bar{m}_{\mu} &= m_{\mu}^*.
 \end{aligned}
\end{equation}

The mathematical equivalence between the hydrodynamic expansion and the near horizon expansion involves a rescaling of the metric as was shown in \cite{Bredberg:2011jq}. In computations we present in the near horizon expansion, we utilize the expansion parameter $\lambda \equiv \frac{\epsilon^2}{r_c}$. Because the $\lambda$ expansion has reorganized the series, we write below the tetrad set used for the near horizon computation, in particular for the type D or rotational velocity and pressure profiles in \eqref{DvsandP}. Note that in this expansion, the coordinates we work with are rescaled to be really $\hat{x}$ and $\hat{y}$; for clarity in the expressions below we have dropped the hats.

The near horizon tetrad we use for the fluid metric dual to \eqref{DvsandP} is
\begin{equation}\label{lambdatetrad}
    \begin{aligned}
    l_{\mu} &= \left\{\frac{1}{\sqrt{2}},0,0,0\right\} + \lambda \left\{\frac{3 r \omega ^2 \left(x^2+y^2\right)}{4 \sqrt{2}},0,0,0\right\} + \lambda^2 \left\{0,-\frac{\omega ^2 \left(x^2+y^2\right)}{2 \sqrt{2}},0,0\right\} \\&\hspace{2cm} + \lambda^3 \left\{\frac{9 r \left(r^2-4\right) \omega ^6 \left(x^2+y^2\right)^3}{64 \sqrt{2}},-\frac{r \omega ^4 \left(x^2+y^2\right)^2}{2 \sqrt{2}},0,0\right\} + \calO(\lambda^4);
\\
    n_{\mu} &= \lambda^{-1} \left\{\frac{r}{\sqrt{2}},0,0,0\right\} + \left\{-\frac{((r-2) r+2) \omega ^2 \left(x^2+y^2\right)}{2 \sqrt{2}},-\sqrt{2},-\frac{r y \omega }{\sqrt{2}},\frac{r x \omega }{\sqrt{2}}\right\}\\& + \lambda \biggr\{\frac{3 r \omega ^4 \left(x^2+y^2\right)^2}{4 \sqrt{2}},\frac{r \omega ^2 \left(x^2+y^2\right)}{\sqrt{2}},-\frac{(r-1) \left(4 q_x+(r-2) y \omega ^3
   \left(x^2+y^2\right)\right)}{2 \sqrt{2}},\\ &\hspace{4cm}\frac{(r-1) \left((r-2) x \omega ^3 \left(x^2+y^2\right)-4 q_y\right)}{2 \sqrt{2}}\biggr\} \\
   &+ \lambda^2\biggr\{0,\frac{\omega  \left(2 (r-2) y g^{(2)}_{rx}-2 (r-2) x g^{(2)}_{ry}+3 (r-1) \omega ^3 \left(x^2+y^2\right)^2\right)}{2 \sqrt{2}},\\&\frac{\omega  \left(4
   (r-2) y g^{(2)}_{xx}+(r-1) \omega  \left(x^2+y^2\right) \left(12 r q_x+(r-1) y \omega ^3 \left(4 (r+1) x^2+(5 r+2) y^2\right)\right)\right)}{8
   \sqrt{2}},\\&\hspace{0.5cm} -\frac{\omega  \left(4 (r-2) x g^{(2)}_{yy}+(r-1) \omega  \left(x^2+y^2\right) \left((r-1) x \omega ^3 \left((5 r+2) x^2+6 r y^2\right)-12 r
   q_y\right)\right)}{8 \sqrt{2}}\biggr\}+\calO(\lambda^3);
\\
   m_{\mu}&= \left\{-\frac{(r-2) \omega  (x-i y)}{2 \sqrt{2}},0,\frac{i}{\sqrt{2}},\frac{1}{\sqrt{2}}\right\} \\& + \lambda\left\{\frac{3 r^2 \omega ^3 (x-i y) \left(x^2+y^2\right)}{8 \sqrt{2}},\frac{\omega  (x-i y)}{\sqrt{2}},\frac{(r-1) y \omega ^2 (x-i y)}{2 \sqrt{2}},-\frac{(r-1) x \omega ^2 (x-i
   y)}{2 \sqrt{2}}\right\}\\&
   + \lambda^2 \biggr\{0,\frac{4 i g^{(2)}_{rx}+4 g^{(2)}_{ry}+(r-2) \omega ^3 (x-i y)^2 (x+i y)}{4 \sqrt{2}},\\&\hspace{1cm}\frac{4 i g^{(2)}_{xx}+(r-1)^2 y \omega ^4 \left(2 x^3-i x^2
   y+2 x y^2-i y^3\right)}{8 \sqrt{2}},\frac{4 g^{(2)}_{yy}-(r-1)^2 x^2 \omega ^4 \left(x^2+y^2\right)}{8 \sqrt{2}}\biggr\} + \calO(\lambda^3);
\\
   \bar{m}_{\mu}& = m_{\mu}^*.
    \end{aligned}
\end{equation}
In the above expressions, the functions $q_i$ and $g_{ij}^{(2)}$ refer to higher order terms necessary in the $\lambda$ expansion to ensure that Einstein's equations are appropriately satisfied, as in \cite{Bredberg:2011jq}.  These functions do not appear in the lowest order Petrov invariants.  Note that this tetrad has been chosen to ensure that $\Psi_2$ is the only nonzero $\Psi_I$; the invariants \eqref{AppendInvar} do not change under tetrad rotations, but the explicit form of the $C_{ABCD}$ in terms of $\iota$ and $o$ does change.  For simplicity, we thus choose a tetrad which preserves \eqref{TypeDWeyl}.

\bibliographystyle{hunsrt}
\bibliography{FluidReferences}
\end{document}